\documentclass[final]{svjour3}
\usepackage[dvipdfmx]{graphicx}
\usepackage{rotating}
\usepackage{amssymb}
\usepackage{mathptmx}
\usepackage[numbers]{natbib}
\makeatletter
\journalname{Journal of Low Temperature Physics}

\usepackage{color}
\usepackage{ulem}
\bibpunct{}{}{,}{s}{}{,}

\begin{document}

\newcommand{\hdblarrow}{H\makebox[0.9ex][l]{$\downdownarrows$}-}
\title{DESHIMA on ASTE: On-sky Responsivity Calibration of the Integrated Superconducting Spectrometer}
\author{Tatsuya Takekoshi$^{1,2,a}$, Kenichi Karatsu$^{3,4}$, Junya Suzuki$^{5,6}$, Yoichi Tamura$^7$, Tai Oshima$^{8,9}$, Akio Taniguchi$^7$, Shin'ichiro Asayama$^8$, Tom J. L. C. Bakx$^{7,8,10}$, Jochem J. A. Baselmans$^{3,4}$, Sjoerd Bosma$^4$, Juan Bueno$^3$, Kah Wuy Chin$^{8,11}$, Yasunori Fujii$^8$, Kazuyuki Fujita$^{12}$, Robert Huiting$^3$, Soh Ikarashi$^4$, Tsuyoshi Ishida$^1$, Shun Ishii$^{8,13}$, Ryohei Kawabe$^{8,9,11}$, Teun M. Klapwijk$^{14,15}$, Kotaro Kohno$^{1,16}$, Akira Kouchi$^{12}$, Nuria Llombart$^4$, Jun Maekawa$^8$, Vignesh Murugesan$^3$, Shunichi Nakatsubo$^{17}$, Masato Naruse$^{18}$, Kazushige Ohtawara$^8$, Alejandro Pascual Laguna$^{3,4}$, Koyo Suzuki$^7$, David J. Thoen$^{4,14}$, Takashi Tsukagoshi$^8$, Tetsutaro Ueda$^7$, Pieter J. de Visser$^3$, Paul P. van der Werf$^{19}$, Stephen J. C. Yates$^{20}$, Yuki Yoshimura$^1$, Ozan Yurduseven$^4$, Akira Endo$^{4,14}$}
\institute{a. \email{tatsuya.takekoshi@ioa.s.u-tokyo.ac.jp}\\
1. Institute of Astronomy, Graduate School of Science, The University of Tokyo, 2-21-1 Osawa, Mitaka, Tokyo 181-0015, Japan \\
2. Graduate School of Informatics and Engineering, The University of Electro-Communications, Cho-fu, Tokyo 182-8585, Japan \\
3. SRON---Netherlands Institute for Space Research, Sorbonnelaan 2, 3584 CA Utrecht, The Netherlands \\
4. Faculty of Electrical Engineering, Mathematics and Computer Science, Delft University of Technology, Mekelweg 4, 2628 CD Delft, the Netherlands \\
5. Department of Physics, Kyoto University, Kyoto 606-8502, Japan \\
6. High Energy Accelerator Research Organization (KEK), 1-1 Oho, Tsukuba, Ibaraki, 305-0801, Japan \\
7. Division of Particle and Astrophysical Science, Graduate School of Science, Nagoya University, Aichi 464-8602, Japan \\
8. National Astronomical Observatory of Japan, Mitaka, Tokyo 181-8588, Japan \\
9. The Graduate University for Advanced Studies (SOKENDAI), 2-21-1 Osawa, Mitaka, Tokyo 181-0015, Japan \\
10 School of Physics \& Astronomy, Cardiff University, The Parade, Cardiff, CF24 3AA, UK \\
11. Department of Astronomy, School of Science, University of Tokyo, Bunkyo, Tokyo, 113-0033, Japan \\
12. Institute of Low Temperature Science, Hokkaido University, Sapporo 060-0819, Japan \\
13. Joint ALMA Observatory, Alonso de C\'ordova 3107, Vitacura, Santiago, Chile \\
14. Kavli Institute of NanoScience, Faculty of Applied Sciences, Delft University of Technology, Lorentzweg 1, 2628 CJ Delft, The Netherlands \\
15. Physics Department, Moscow State Pedagogical University, 119991 Moscow, Russia \\
16. Research Center for the Early Universe, Graduate School of Science, The University of Tokyo, 7-3-1 Hongo, Bunkyo-ku, Tokyo 113-0033, Japan \\
17. Institute of Space and Astronautical Science, Japan Aerospace Exploration Agency, Sagamihara 252-5210, Japan \\
18. Graduate School of Science and Engineering, Saitama University, 255, Shimo-okubo, Sakura, Saitama 338-8570, Japan \\
19. Leiden Observatory, Leiden University, PO Box 9513, NL-2300 RA Leiden, The Netherlands \\
20. SRON, Landleven 12, 9747 AD Groningen, The Netherlands}

\authorrunning{Takekoshi {\it et al.}}
\titlerunning{Responsivity Calibration of DESHIMA}

\maketitle

\begin{abstract}
We are developing an ultra-wideband spectroscopic instrument, DESHIMA (DEep Spectroscopic HIgh-redshift MApper), based on the technologies of an on-chip filter-bank and Microwave Kinetic Inductance Detector (MKID) to investigate dusty star-burst galaxies in the distant universe at millimeter and submillimeter wavelength.
An on-site experiment of DESHIMA was performed using the ASTE 10-m telescope.
We established a responsivity model that converts frequency responses of the MKIDs to line-of-sight brightness temperature.
We estimated two parameters of the responsivity model using a set of skydip data taken under various precipitable water vapor (PWV, 0.4--3.0~mm) conditions for each MKID.
The line-of-sight brightness temperature of sky is estimated using an atmospheric transmission model and the PWVs.
As a result, we obtain an average temperature calibration uncertainty of $1\sigma=4$\%, which is smaller than other photometric biases.
In addition, the average forward efficiency of 0.88 in our responsivity model is consistent with the value expected from the geometrical support structure of the telescope.
We also estimate line-of-sight PWVs of each skydip observation using the frequency response of MKIDs, and confirm the consistency with PWVs reported by the Atacama Large Millimeter/submillimeter Array.

\end{abstract}
\section{Introduction}
Wideband spectroscopic measurement at millimeter/submillimeter wavelengths with moderate spectral resolutions (resolving power $\nu/\Delta \nu\sim 1000$) is an important astronomical approach to obtain physical and chemical properties of the cold gas component in galaxies and their spectroscopic redshifts in the early universe via emission from ionized atoms (e.g., [C{\small II}] 158~$\mathrm{\mu m}$, [O{\small III}] 88~$\mathrm{\mu m}$) and molecules (e.g., CO, H$_2$O)\cite{Stacey2011}. 
To efficiently detect astronomical line emission from galaxies, we are developing a new spectrometer, DESHIMA (DEep Spectroscopic HIgh-redshift MApper), based on an on-chip superconducting filter-bank technology integrated with microwave kinetic inductance detectors (MKIDs)\cite{2019JATIS...5c5004E}.
In October and November, 2017, we had a first on-sky demonstration of the instrument, covering 332--377~GHz with 49 spectral channels ($\nu/\Delta \nu\sim 380$) using the Atacama Submillimeter Telescope Experiment 10-m telescope (ASTE) operated in the Atacama desert, Chile\cite{2004SPIE.5489..763E, ezawa2008new}.
In this session, we successfully detected some astronomical targets, for instance, a redshifted CO($J=$3--2) line of VV~114, a luminous infrared galaxy at the redshift of 0.020\cite{2019NatAs.tmp..418E}.

Establishing an on-sky responsivity model is a key issue in realizing astronomical observations using the wideband spectroscopic instrument with the newly-deployed MKID technology and readout system\cite{2017A&A...601A..89B}.
The responsivity model converts raw readout values of MKIDs to line-of-sight brightness temperatures of the sky under various atmospheric conditions by considering optical loading from the telescope and warm optics.
The responsivity model parameters of MKIDs on-sky are assumed to be different from those obtained in the laboratory because the 4 K stage temperature and magnetic field environment are different.
Thus, it is important to establish a reliable responsivity calibration method that estimates the responsivity model parameters of MKIDs and optical loading from warm optics simultaneously.

In this paper, we present a new calibration scheme used for the astronomical data analysis of the first on-site DESHIMA session\cite{2019NatAs.tmp..418E}.

\section{Method}

In this study, we aim to establish the on-sky responsivity model between the line-of-sight brightness temperature of the sky, $T_\mathrm{sky}$, and the resonance frequency shift of a MKID in our readout system\cite{2017A&A...601A..89B}.
Before the on-site session using the ASTE telescope, we established the responsivity model using a cold load at 77--300~K outside the cryostat window in the laboratory measurement\cite{2019JATIS...5c5004E}.
Using the readout resonance frequency of $f (T)$, which depends on the optical loading to the MKIDs, we define the relative frequency shift: $\delta x (T) \equiv (f(T)- f_\mathrm{load})/ f_\mathrm{load} $, where $T$ is the Rayleigh-Jeans brightness temperature outside the cryostat window and $f_\mathrm{load}$ is the resonance frequency when using the room-temperature thermal load mounted outside the cryostat window.
Using the input optical power $P$ dependency of NbTiN-Al hybrid MKID response ($\delta x \propto \sqrt{P}$)\cite{2019JATIS...5c5004E,2014NatCo...5.3130D,2013ApPhL.103t3503J, 2014ApPhL.105s3504J}, the relation between $\delta x$ and $T$ are written as
\begin{equation}
\label{eq:1}
\delta x (T) = p_0 (\sqrt{T + T_\mathrm{0}} - \sqrt{T_\mathrm{load} + T_\mathrm{0}} ),
\end{equation}
where $p_0$ is a factor of proportionality depending on the sensitivity of the MKIDs, $T_\mathrm{0} = -\frac{h\nu}{2k_b}$ ($=-8.4$~K at 350~GHz) is a constant correction temperature derived as the second order expansion term of the Planck function, and $T_\mathrm{load}$ is the brightness temperature of the room-temperature thermal load.
This linearized approximation of the Planck function is more accurate (-0.03\% offset from the Planck function) than the Rayleigh-Jeans (+2.8\%) approximation for 300~K physical temperature emission at 350~GHz, and reduces the computational cost in our calibration scheme.

Here, we would like to model the relation between $\delta x$ and the line-of-sight brightness temperature purely coupling to the sky, $T_\mathrm{sky}$.
Thus, it is necessary to consider the spillover radiation load from the warm telescope optics.
We can define the brightness temperature outside the cryostat window $T = \eta_\mathrm{fwd} T_\mathrm{sky} + (1-\eta_\mathrm{fwd}) T_\mathrm{amb}$, where $\eta_\mathrm{fwd}$ is the forward efficiency of the telescope and warm optics, and $T_\mathrm{amb}$ is the ambient temperature of the telescope.
Solving with Eq. (\ref{eq:1}), the sky temperature response can be written as
\begin{equation}
\label{eq:2}
T^\mathrm{resp}_\mathrm{sky} (\delta x) = \frac{1}{p_0^2 \eta_\mathrm{fwd}}(\delta x + p_0\sqrt{T_\mathrm{load}+T_\mathrm{0}})^2-\frac{1}{\eta_\mathrm{fwd}}(T_\mathrm{0}+(1-\eta_\mathrm{fwd})T_\mathrm{amb}).
\end{equation}
Just before the observations, our readout system measures $f_\mathrm{load}$ under the optical load from a room-temperature blackbody chopper mounted outside the cryostat window\cite{2019NatAs.tmp..418E}. 
This system was developed based on the multi-temperature calibrator system for the multi-chroic camera system for the ASTE telescope\cite{2018JLTP..193.1003T}, and used a mixture of SiC grains and carbon loaded epoxy as blackbody\cite{2019JATIS...5c5004E,klaassen2001optical}.
We take an effective emissivity $\kappa_\mathrm{eff}=0.95$ for the blackbody chopper, which is estimated from an emissivity measurement using the 350~GHz transition edge sensor camera system that we developed previously\cite{2018JLTP..193.996O, hirota2013development, oshima2013development}.
We assume that the reflected portion of the beam is coupled to the cold interior of the cryostat.
Thus, we use $T_\mathrm{load} = \kappa_\mathrm{eff} T_\mathrm{room}$, where $T_\mathrm{room} \simeq 289$~K is the room temperature in the ASTE cabin. 
We also use the measured outside temperature at the ASTE site as $T_\mathrm{amb}$.

To determine the responsivity model parameters, $p_0$ and $\eta_\mathrm{fwd}$, in Eq. (\ref{eq:2}), we compare observed $\delta x$ and  $T_\mathrm{sky}$, which is estimated from the atmospheric conditions and telescope elevation angle $El$.
Considering the frequency dependence of optical transmission of each filter pixel, $g(\nu)$, we can define atmospheric transmission, 
\begin{equation}
\label{eq:3}
\eta_\mathrm{atm} (El, PWV) =\frac{\int g(\nu) e^{-\tau_0(\nu, PWV)\csc(El) }\mathrm{d}\nu}{\int g(\nu)\mathrm{d}\nu},
\end{equation}
where $\tau_0$ is the zenith opacity, which depends on precipitable water vapor ($PWV$), and $\csc(El)$ is the airmass assuming the plane-parallel approximation of atmospheric structure.
The relation between $\tau_0$ and $PWV$ is estimated by a radiative transfer atmospheric model, ATM\cite{2001ITAP...49.1683P}.
By ignoring the contribution of the cosmic microwave background radiation because it is in the Wien region of the Planck radiation, the estimated line-of-sight brightness temperature of the sky at the elevation angle $El$ can be estimated using the equation:
\begin{equation}
\label{eq:4}
T^\mathrm{los}_\mathrm{sky} (El, PWV) = (1-\eta_\mathrm{atm})T_\mathrm{atm},
\end{equation}
where $T_\mathrm{atm}$ is the atmospheric temperature.
We assume that $T_\mathrm{atm}=T_\mathrm{amb}$.

During the on-site DESHIMA session, we executed 22 skydip measurements, each of which was composed of two round-trips in the elevation angle range of $32^\circ$ to $88^\circ$ in $\sim$2 minutes.
Using these  skydip data sets, we established the responsivity models for each MKID and estimated PWVs using DESHIMA itself in the following manner.
First, we made an initial estimation of $T^\mathrm{los}_\mathrm{sky}(El, PWV)$ using Eq.~(\ref{eq:4}) and the PWVs  (0.4--3.0~mm, median 0.9~mm) obtained by the radiometers of the Atacama Large Millimeter/submillimeter Array (hereafter ALMA PWVs)\cite{2013A&A...552A.104N}.
Assuming $T^\mathrm{resp}_\mathrm{sky}=T^\mathrm{los}_\mathrm{sky}$, we fit the parameters of the responsivity model ($p_0$ and $\eta_\mathrm{fwd}$) using Eq.~(\ref{eq:2}) for each MKID channel.
Then, we re-estimated PWVs using the obtained responsivity model of MKIDs (DESHIMA PWVs), by comparing the obtained spectrum using Eq.~(\ref{eq:2}), $T^\mathrm{resp}_\mathrm{sky}$, with the atmospheric transmission model. 
As can be seen in Fig.~\ref{fig:calcurve}, the systematic difference between $T^\mathrm{resp}_\mathrm{sky}$ and $T^\mathrm{los}_\mathrm{sky}$ become smaller if we adopt the DESHIMA PWV, compared to the initial ALMA PWV.
This is most likely because the DESHMA PWV is based on a real-time line-of-sight measurement by DESHIMA on ASTE, whilst the ALMA PWV represents the PWV above the center of the ALMA antennas located $\sim8$~km away from the ASTE telescope.
We do not re-estimate the responsivity model parameters using the DESHIMA PWV to avoid error propagation, although further iterative estimates of the responsivity model parameters could potentially improve the model.

\section{Result}
\begin{figure}
\begin{center}
\includegraphics[width=0.9\linewidth, keepaspectratio]{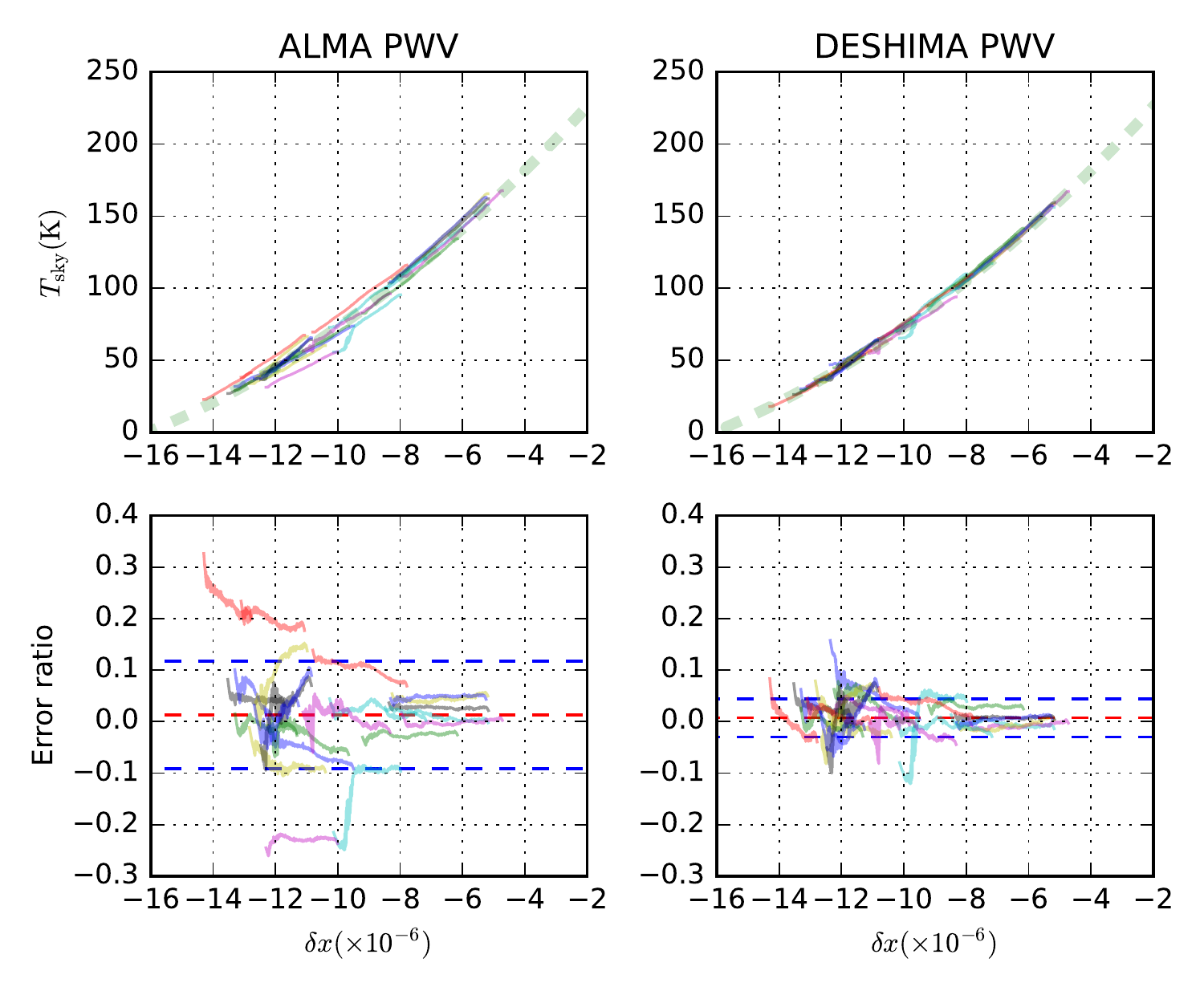}
\caption{An example of the responsivity model fit  (KID \#20, 335.5~GHz). \textit{Top Left} and \textit{Right} panels show the relation between $\delta x$ and $T_\mathrm{sky}$ calculated using ALMA and DESHIMA PWVs, respectively. \textit{Dashed green lines} represent the obtained responsively model curve of $T^\mathrm{resp}_\mathrm{sky}(\delta x)$. \textit{Solid lines} correspond to the relations between $T^\mathrm{los}_\mathrm{sky}$ and $\delta x$ of each skydip data. \textit{Bottom Left} and \textit{Right} panels show the error ratio of responsivity curve. \textit{Red and blue dashed lines} corresponds to the mean and standard deviation of the error ratio. (Color figure online.) 
\label{fig:calcurve}}
\end{center}
\end{figure}

\begin{figure}
\begin{center}
\includegraphics[width=0.9\linewidth, keepaspectratio]{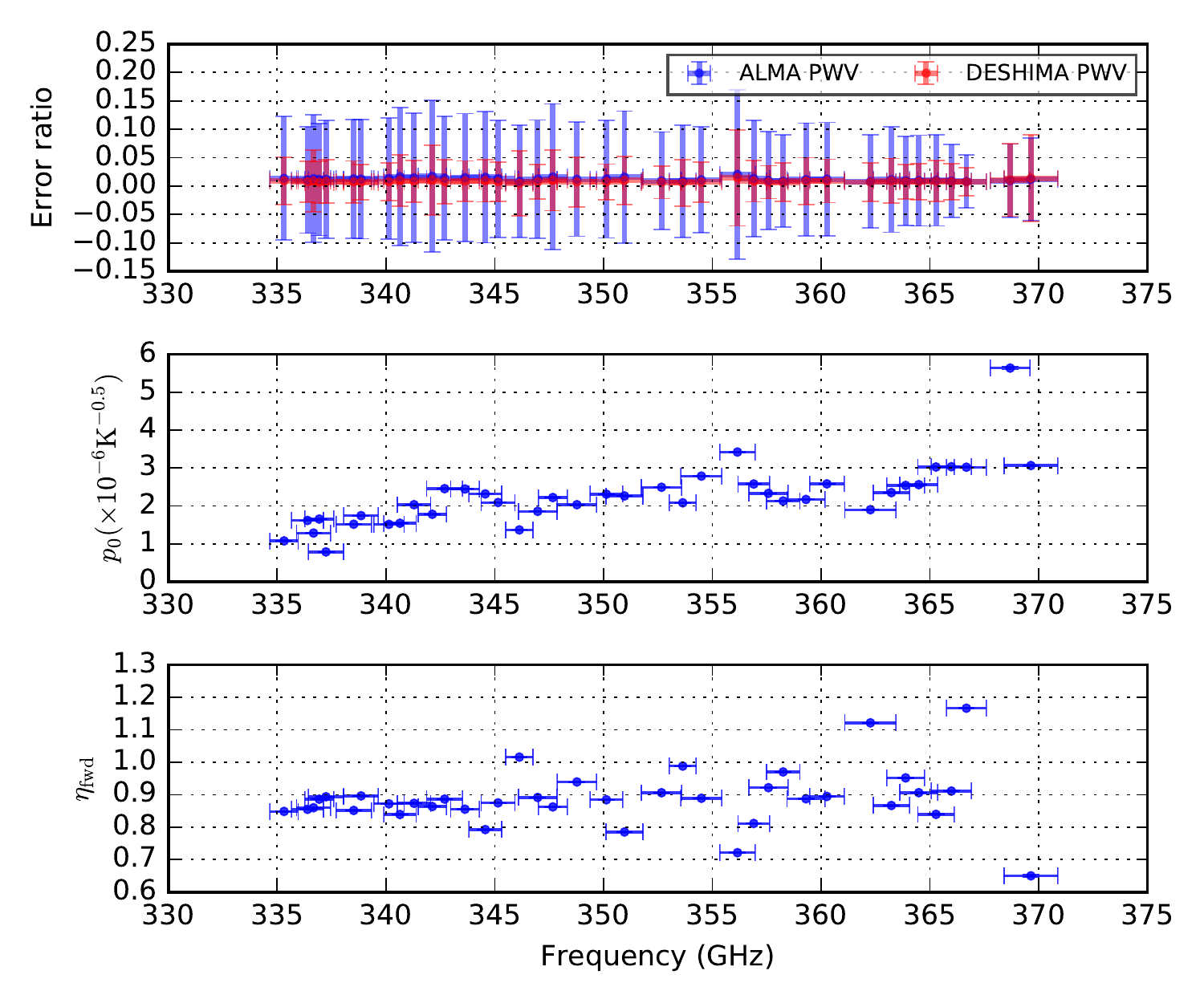}
\caption{Frequency dependencies of the error ratio (\textit{Top}) and responsivity model parameters, $p_0$ (\textit{Middle}) and $\eta_\mathrm{fwd}$ (\textit{Bottom}). Horizontal and vertical error bars show the effective bandwidths and the standard deviations of the error ratios, respectively. (Color figure online.) 
\label{fig:calparams}}
\end{center}
\end{figure}

We obtained the responsivity model parameters for 40 filter channels. 
An example of the responsivity model fit for the 20th channel, corresponding to a filter frequency of 335.5~GHz, is shown in Fig.~\ref{fig:calcurve}.
The relations between $\delta x$ and $T^\mathrm{los}_\mathrm{sky}$ were explained well by the obtained responsivity model curves. 
We defined an error ratio as $(T^\mathrm{los}_\mathrm{sky}-T^\mathrm{resp}_\mathrm{sky})/T^\mathrm{resp}_\mathrm{sky}$ to check the improvement using the DESHIMA PWVs.
Fig.~\ref{fig:calparams} shows the means and standard deviations of the error ratios for each filter channels, which correspond to the systematic offset and $1\sigma$ uncertainty of the sky temperature estimate.
As the statistics of the 40 filter channels, the mean value of systematic offsets and standard deviations were 1.0\% and 9.9\% using the ALMA PWVs, respectively, and 0.8\% and 4.0\% using the DESHIMA PWV, respectively.
Thus, we confirmed the improvement of the standard deviation using the DESHIMA PWVs.
The systematic offset can be removed by multiplying the factor to the obtained $T_\mathrm{sky}$.
The standard deviation is expected to arise mainly from atmospheric fluctuation in time and space distribution, and contributes to the photometric error of brightness temperatures.
However, the obtained standard deviations are much smaller than other systematic uncertainties in intensity scaling, such as the chopper wheel calibration method and the planet flux model we employed (typically 5--20\%).
Thus, this would not be a dominant component of photometric errors.

Fig.~\ref{fig:calparams} also shows the frequency dependencies of the responsivity model parameters. 
The averages (standard deviations) of $p_0$ and the forward efficiency were $2.2(\pm 0.8)\times10^{-6} \mathrm{K^{-0.5}}$ and 0.88($\pm$0.12), respectively. 
The estimated forward efficiency is consistent with the efficiency expected from aperture blockage caused by the geometry of the sub-reflector and its support legs of the ASTE telescope, $\eta^\mathrm{ASTE}_\mathrm{fwd}=0.87$.
The offsets from $\eta^\mathrm{ASTE}_\mathrm{fwd}$ in many channels could be attributed to additional offset of input power to the detector, for example, optical coupling to the cold stages in the cryostat, or spillover of the warm optics components caused by optical misalignment between cold and warm optics\cite{2019JATIS...5c5004E,2019NatAs.tmp..418E}.

\begin{figure}
\begin{center}
\includegraphics[width=0.5\linewidth, keepaspectratio]{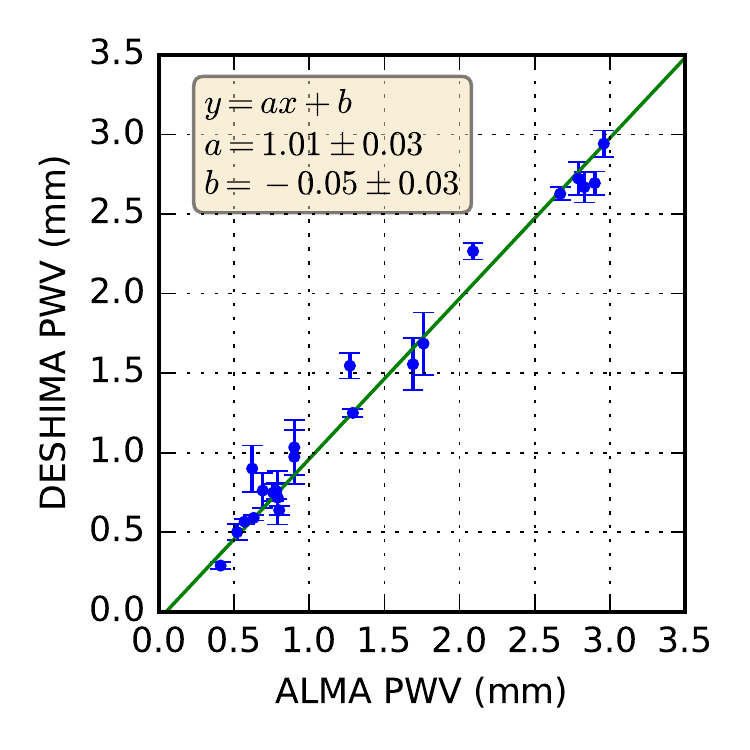}
\caption{Relation between ALMA and DESHIMA PWVs. \textit{Green line} shows the best-fit linear function of the relation. (Color figure online.) 
\label{fig:PWV}}
\end{center}
\end{figure}

We also checked the consistency between the ALMA and DESHIMA PWVs.
As shown in Fig.~\ref{fig:PWV}, we confirmed a linear relationship between the ALMA and DESHIMA PWVs.
This supports that the DESHIMA PWVs using the obtained responsivity parameters are reliable and allow us to use DESHIMA as a line-of-sight radiometer for the ASTE telescope.

\section{Conclusion}
We reported an on-sky responsivity calibration method for DESHIMA using the skydip data. 
Our responsivity model reproduces the expected forward efficiency of the ASTE telescope, and archives a photometric error of $1\sigma =4$\%, which is better than other photometric errors.
We are planning for the next DESHIMA session to use a new filter-bank chip operated at 220--440~GHz with $\nu/\Delta\nu \sim 500$.
Our responsivity calibration method will be tested for further improvement in the next on-site operation.
In particular, on-site, line-of-sight, and high-frequency water vapor radiometer system refined from an opacity imager operated at 183~GHz\cite{2011PASJ...63..347T} will help to consider atmospheric fluctuation and improve the photometric accuracy of the spectrometer.
Our method can also be used for other ultra-wideband spectroscopic instrument using MKIDs such as SuperSpec\cite{Shirokoff2014} and Micro-Spec\cite{Cataldo2018}.

\begin{acknowledgements}
This research was supported by the Netherlands Organization for Scientific Research NWO (Vidi grant No. 639.042.423, NWO Medium Investment grant No. 614.061.611 DESHIMA), the European Research Council ERC (ERC-CoG-2014 - Proposal n$^\circ$ 648135 MOSAIC), the Japan Society for the Promotion of Science JSPS (KAKENHI Grant Numbers JP25247019, JP17H06130 and JP19K14754), and the Grant for Joint Research Program of the Institute of Low Temperature Science, Hokkaido University. Y.T. and T.J.L.C.B. are supported by NAOJ ALMA Scientific Research Grant Number 2018-09B. P.J. de V. is supported by the NWO (Veni Grant 639.041.750). T.M.K. is supported by the ERC Advanced Grant No. 339306 (METIQUM) and the Russian Science Foundation (Grant No. 17-72-30036). N.L. is supported by ERC (Starting Grant No. 639749). J.S. and M.N. are supported by the JSPS Program for Advancing Strategic International Networks to Accelerate the Circulation of Talented Researchers (Program No. R2804). T.J.L.C.B was supported by the European Union Seventh Framework Programme (FP7/2007--2013, FP7/2007--2011) under grant agreement No. 607254. Data analysis was, in part, carried out on the open-use data analysis computer system at the Astronomy Data Center of NAOJ. The ASTE telescope is operated by National Astronomical Observatory of Japan (NAOJ).
\end{acknowledgements}

\end{document}